\documentclass[preprint,groupedaddress,nofootinbib,preprintnumbers,eqsecnum]{revtex4-1}

\usepackage{amsmath}
\usepackage{mathrsfs}
\usepackage[psamsfonts]{amssymb}
\usepackage{bm} 
\usepackage[dvipdfmx]{hyperref}
\begin{document}


\title{Gaugeon formalism for the two-form gauge fields}


\author{Masataka Aochi}
\affiliation{Asahikawa Jitsugyo High School, Asahikawa, Hokkaido 071-8138, Japan}
\author{Ryusuke Endo}
\email{endo@sci.kj.yamagata-u.ac.jp}-
\affiliation{Department of Physics, Yamagata University, Yamagata, 990-8560, Japan}
\author{Hikaru Miura}
\affiliation{Yamagata Meisei High School, Yamagata 990-2322, Japan}

%
\def\brst{\bm{\delta}_\mathrm{B}}
\def\tbrst{\tilde{\bm{\delta}}_\mathrm{B}}
%
\begin{abstract}
We present a BRST symmetric gaugeon formalism for the two-form gauge fields. 
A set of vector gaugeon fields is introduced as a quantum gauge freedom. 
One of the gaugeon fields satisfies a higher derivative field equation; 
this property is necessary to change the gauge-fixing parameter of the two-form 
gauge field. 
A naive Lagrangian for the vector gaugeon fields is itself invariant 
under a gauge transformation for the vector gaugeon field. 
The Lagrangian of our theory includes 
the gauge-fixing terms for the gaugeon fields and 
corresponding Faddeev--Popov ghosts terms. 
\end{abstract}

\preprint{YGHP-17-09}

\pacs{}

\maketitle

%
%
%
\section{Introduction}
%
The standard formalism of canonically quantized gauge 
theories \cite{Nakanishi,Kugo78a,Kugo78b,Kugo79,Kugo89} 
does not consider quantum-level gauge transformations. 
There is no quantum gauge freedom, 
since the quantum theory is defined only after the gauge fixing.
Within the broader framework of Yokoyama's gaugeon 
formalism \cite{Yokoyama74a}, 
we  can consider quantum gauge transformations as $q$-number gauge transformations. 
In this formalism, quantum gauge freedom is provided 
by an  extra field, called a gaugeon field. 
The gaugeon formalism has been developed  so far for various gauge fields, such as, 
Abelian gauge fields \cite{Yokoyama74a,Yokoyama74b,Izawa,Koseki,Endo,Saito}, 
non-Abelian gauge fields \cite{Yokoyama78b,Yokoyama78c,Yokoyama78d,Yokoyama78e,
Yokoyama80,Abe,Koseki96,Sakoda}, 
Higgs models \cite{Yokoyama75a,Miura}, 
chiral gauge theories \cite{Yokoyama75b}, 
Schwinger's model \cite{Nakawaki}, 
spin-3/2 gauge fields \cite{Endo00}, 
string theories \cite{Faizal12a,Faizal12b}, 
and gravitational fields \cite{Upadhyay14a,Upadhyay14b}. 

Recently, gaugeon formalisms for the Abelian two-form gauge fields are 
considered  
by Upadhyay and Panigrahi \cite{UpadhyayPanigrahi} 
(in the framework of the ``very special relativity" \cite{Glashow}), 
and by Dwivedi \cite{Dwivedi}. 
They introduced a vector gaugeon field which would play a role of the quantum gauge 
freedom of the two-form gauge field. 
The vector gaugeon field itself has a property of  gauge fields. It has a 
gauge invariance. 
In fact, the Lagrangians given in Refs.\cite{UpadhyayPanigrahi,Dwivedi} 
are invariant under the gauge transformation of the vector gaugeon field. 
So, we should fix the gauge before quantizing the vector gaugeon field. 
However, the authors of Refs.\cite{UpadhyayPanigrahi,Dwivedi} 
did not fix the gauge. Thus, their vector gaugeon field was not quantized. 
Namely, their theories are incomplete as a gaugeon 
formalism for the two-form gauge fields; 
they do not permit the quantum level gauge transformation, 
which is an essential ingredient of the gaugeon formalism. 

The aim of this paper is quantizing the vector gaugeon field 
and obtaining a correct gaugeon 
theory for the two-form gauge field.

This paper is organized as follows. In sect. 2, we first review the standard formalism 
for the covariantly quantized two-form gauge field. Then, we show that 
the vector gaugeon field must be a massless dipole field, that is, 
its propagator have a term proportional to $1/(p^2)^2$.
In sect. 3, we covariantly fix the gauge of the massless dipole vector field 
and quantize the system. 
In section 4, incorporating the massless dipole vector field as the gaugeon field, 
we present a correct gaugeon theory of the two-form gauge field. 
Section 5 is devoted to summary and comments. 

%
%
\section{standard formalism}
%
A Faddeev-Popov quantization of the antisymmetric tensor gauge field 
(the two-form gauge field) \cite{Hayashi, KalbRamond}   
was first performed by Townsend \cite{Townsend}. 
He revealed that the Faddeev-Popov (FP) 
ghosts themselves have  gauge invariance and thus 
the ghosts for ghosts are necessary. 
His theory, however, violates unitarity 
because of inappropriate ghost contents. 
To ensure the unitarity, counting of ghosts should have been improved. 
The correct mode-counting was given  by Kimura \cite{Kimura2form} 
and Siegel \cite{Siegel}.
In the BRST quantization scheme \cite{Kugo78a,Kugo78b,Kugo79}, 
Kimura \cite{Kimura2form} has  introduced a correct number of 
FP ghosts and auxiliary multiplier fields which form an off-shell nilpotent 
BRST symmetry. The unitarity of the theory is assured by Kugo-Ojima's mechanism 
of BRST quartets \cite{Kugo79,Kugo89}. Kimura also gave canonically  quantized theories 
of the antisymmetric tensor gauge fields of third rank \cite{Kimura3form}  and 
of arbitrary rank \cite{KimuraArbitrary}.%
            \footnote{Kimura's Lagrangians were also given 
            \cite{Tonin, Kawasaki,Thierry-Mieg}  
                      by Bonora-Tonin's superspace 
                      method \cite{BonoraTonin} of the BRST symmetry. 
                      } 
In the path integral formalism, Siegel \cite{Siegel} gave the  precise ghost counting 
by a careful application of the 't Hooft averaging to the arbitrary rank 
antisymmetric tensor gauge fields.\footnote{%
           See also 
           Ref. \cite{Obukhov,Obukhov2}. 
           } 
In this section, we review  Kimura's theory as a standard formalism. 

The classical (gauge-unfixed) Lagrangian of a two-form 
gauge field $B_{\mu\nu}$ 
is given by
\begin{align}
  \mathcal{L}_0= \frac{1}{12} F^{\lambda\mu\nu}F_{\lambda\mu\nu},
  \label{L:classical}
\end{align}  
where the third-rank antisymmetric tensor $F_{\lambda\mu\nu}$ is the field strength 
of $B_{\mu\nu}$ defined by 
\begin{align}
    F_{\lambda\mu\nu}=  \partial_\lambda B_{\mu\nu}
                      + \partial_\mu B_{\nu\lambda}
                      + \partial_\nu B_{\lambda\mu}.
\end{align}
The tensor 
$F_{\lambda\mu\nu}$
and thus the Lagrangian (\ref{L:classical}) are invariant 
under the gauge transformation
\begin{align}
  B_{\mu\nu} \to 
                 B_{\mu\nu}
                              +  
                                   \partial_\mu \Lambda_\nu
                                 -  \partial_\nu \Lambda_\mu
                                 ,
  \label{GaugeTransformationOfBmn}                                 
\end{align}
where $\Lambda_\mu$ is an arbitrary vector field. Thus, to obtain a quantized 
theory, we need gauge-fixing and 
appropriate ghosts and auxiliary fields. 
Note that 
the second term on the right hand side of 
(\ref{GaugeTransformationOfBmn}) 
is invariant under a ``gauge transformation" 
$\Lambda_\mu \to \Lambda_\mu + \partial_\mu \Lambda$ with an arbitrary scalar 
function $\Lambda$. This is the origin why we need ghosts for ghosts in the 
quantized theory of the antisymmetric tensor gauge theories. 

\begin{figure}
  \centering
  \setlength{\unitlength}{0.6mm}
  \begin{picture}(80,99)(25,-3)
    {\thinlines 
      \put(90,-3){\vector(0,1){93}}
         \put(93,90){ghost \#} 
          \put(94,79){$2$}    \put(89,82){\line(1,0){2}}
          \put(94,60){$1$}    \put(89,62){\line(1,0){2}}
          \put(94,40){$0$}    \put(89,42){\line(1,0){2}}
          \put(93,20){$-1$}   \put(89,22){\line(1,0){2}}
          \put(93,0){$-2$}    \put(89, 2){\line(1,0){2}}
      }
     \put(10,40){$B_{\mu\nu}$}
          \put(32,40){$B_\mu$}
                 \put(54,40){$\eta$}
          \put(32,60){$c_\mu$}
          \put(32,20){$c_{\ast \mu}$}
                 \put(54,60){$d$}
                 \put(54,20){$d_{\ast}$}
                 \put(54,80){$\phi$}
                 \put(54,0){$\phi_{\ast}$}
     \put(19,46){\vector(1,1){11}}
     \put(41,66){\vector(1,1){11}}
          \put(35,26){\vector(0,1){10}}
                 \put(56,46){\vector(0,1){10}}
                 \put(56, 6){\vector(0,1){10}}
 \end{picture}
 \caption{
         Field contents and  BRST transformations of Kimura's theory. 
         The arrows represent the directions of the BRST transformations. 
         The fields of odd ghost numbers are fermionic, while those of 
         even ghosts numbers  bosonic.
         }\label{fig:1}
\end{figure}
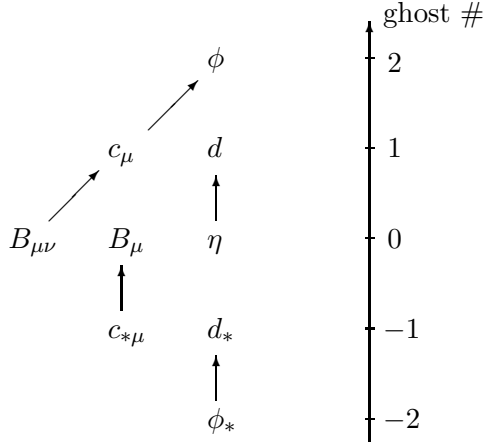

The quantum Lagrangian given by Kimura \cite{Kimura2form} is
\begin{align}
   \mathcal{L}_\mathrm{K}=&\mathcal{L}_0  
                -\partial ^\mu B^\nu B_{\mu\nu}
               - \frac{\alpha}{2} B^\mu B_\mu 
               + B^\mu \partial_\mu \eta 
              + \partial^\mu \phi_\ast \partial_\mu \phi
            \notag \\ 
            &
              -\frac{i}{2}(\partial^\mu c_{\ast}^{\, \nu}-\partial^\nu c_{\ast}^{\, \mu})
                           (\partial_\mu c_\nu -\partial _\nu c_\mu)
            +i c_{\ast}^{\, \mu} \partial _\mu d 
            +i\partial^\mu d_\ast c_\mu 
            + i\beta d_\ast d ,
\label{L:Kimura}
\end{align}
where $\alpha$ and $\beta$ are real parameters, 
$B_\mu$ is  (partly) a multiplier field imposing a gauge condition 
$\partial^\mu B_{\mu\nu}= \alpha B_\nu + \cdots$ on $B_{\mu\nu}$ as a field equation, 
$c_\mu$ and $c_{\ast \mu}$ are FP ghosts, and  scalar fields 
$\phi$, $\phi_\ast$, $d$, $d_\ast$ and $\eta$ play the roles of 
ghosts for ghosts or multiplier fields. 
One may expect 
these roles by observing the 
following BRST transformations under which 
Kimura's  Lagrangian (\ref{L:Kimura}) is invariant:\footnote{%
          The Lagrangian   (\ref{L:Kimura})  is also invariant under 
          the anti-BRST transformation \cite{Tonin, Kawasaki, Thierry-Mieg}. 
          }%
\begin{align}
  &\brst B_{\mu\nu}= \partial_\mu c_\nu -  \partial_\nu c_\mu,
    \qquad 
    \brst c_\mu= -i \partial_\mu  \phi, 
    \notag \\
  & \brst c_{\ast \mu}= iB_\mu, 
      \qquad 
        \brst \phi_\ast = d_\ast, 
          \qquad
             \brst \eta = d, 
    \label{BRST:Kimura}
    \\%
  & \brst B_\mu= \brst d_\ast =\brst d = \brst \phi=0. 
  \notag
\end{align}
These  BRST transformations 
satisfy the off-shell nilpotency  $\brst ^2=0$. 
The corresponding  BRST charge $Q_{\rm B(K)}$ 
can be written as 
\begin{align}
  Q_\mathrm{B(K)} = \int \left[
                    B^\lambda \overleftrightarrow{\partial_0} c_\lambda
                   + d_\ast \overleftrightarrow{\partial_0} \phi
                   + (1-\beta) B_0 d
                   \right] d^{D-1} x ,
                  \label{QB:Kimura}
\end{align}
where we consider in $D$-dimensional space-time, 
and 
$
\overleftrightarrow {\partial_0} = \overrightarrow{\partial_0}
                                   - \overleftarrow{\partial_0}
$. 
This charge  is also nilpotent:  
$Q_\mathrm{B(K)}^{\,2}=0$. 
Figure \ref{fig:1} shows the field contents and their BRST transformations. 
%
With these field contents, Kugo--Ojima's 
quartet mechanism \cite{Kugo79,Kugo89} works 
and all the unphysical modes are removed by Kugo--Ojima's physical subsidiary 
condition, 
\begin{align}
     Q_\mathrm{B(K)} |\mathrm{phys}\rangle=0. 
\end{align}
Especially, 
the fields $\eta$ and $d$ are necessary in correct mode-counting; 
without these fields 
the longitudinal modes of $B_\mu$ and $c_{\ast \mu}$ 
could not form a BRST quartet. 
 
The field equations for the zero-ghost-number fields  derived from (\ref{L:Kimura}) 
are 
\begin{align}
   & \partial^\lambda F_{\lambda\mu\nu}
                    + \partial _\mu B_\nu - \partial _\nu B_\mu =0 , 
   \\
   & \partial^\lambda B_{\lambda \mu} - \alpha B_\mu + \partial_\mu \eta =0,
                                \label{Eq:gaugecondition}
   \\
   & \partial^\lambda B_\lambda =0, 
                                 \label{Eq:BoxB}
\intertext{from which we also have}
 &  \Box B_\mu =  \Box \eta =0.
\end{align}

We regard the equation (\ref{Eq:gaugecondition}) as the Lorenz-like 
gauge condition 
for the gauge field $B_{\mu\nu}$ and  $\alpha$ as a gauge-fixing parameter. 
Now we consider a possibility to change the gauge-fixing parameter $\alpha$ 
by an appropriate $q$-number gauge transformation, which would be 
given by 
\begin{align} 
  B_{\mu\nu} \to \hat B_{\mu\nu}=B_{\mu\nu}
              + \tau (\partial_\mu Y_\nu - \partial_\nu Y_\mu), 
\label{Poss:qtransf}
\end{align}
where the vector field $Y_\mu$ is a would-be gaugeon field and $\tau$ is a
real parameter.  
One possibility  is that 
$Y_\mu$ satisfies
\begin{align}
   \partial^\mu (\partial_\mu Y_\nu - \partial_\nu Y_\mu ) = B_\nu, 
   \label{Poss:eq.forY}
\end{align}
so that the gauge condition (\ref{Eq:gaugecondition}) transforms 
under (\ref{Poss:qtransf}) as 
\begin{align}
   \partial ^\mu \hat B_{\mu \nu} - (\alpha + \tau) B_\nu + \partial_\nu \eta=0. 
\end{align}
Thus the gauge-fixing parameter changes from $\alpha$ to $\alpha + \tau$. 
From (\ref{Poss:eq.forY}) together with (\ref{Eq:BoxB}) 
we presume the field equation 
for the gaugeon field $Y_\mu$ to be 
\begin{align}
    \Box \partial^\mu (\partial_\mu Y_\nu - \partial_\nu Y_\mu )=0, 
                \label{Eq:DipoleVector}
\end{align}
which suggests that the gaugeon for the two-form $B_{\mu\nu}$ 
 would be a  massless dipole field.

\section{quantum theory of a massless dipole vector field}
%
\subsection{classical theory}
Here we consider the quantization of the massless dipole vector field $Y_\mu$, 
whose classical equation is given by (\ref{Eq:DipoleVector}). 
To avoid a higher derivative Lagrangian we imitate the Froissart model 
\cite{Froissart} describing a dipole scalar field. 
Simply generalizing the Froissart model to our case, we adopt 
\begin{align}  
   \mathcal{L}_\mathrm{vF0}
      = -\frac{1}{2} (\partial^\mu Y_\ast^\nu - \partial^\nu Y_\ast^\mu )
                    (\partial_\mu Y_\nu - \partial_\nu Y_\mu )
        - \frac{\varepsilon}{2} Y_\ast^\mu Y_{\ast\mu}
                \label{L:vectorFroissart}
\end{align}
as a starting Lagrangian, where $\varepsilon$ is a 
sign factor $\varepsilon = \pm 1$, and $Y_{\ast \mu}$ is an auxiliary vector 
field. We call this model 
a massless vector-Froissart model.\footnote{%
       A brief report of the quantization of this model 
       was given by one of the authors (M. A.) \cite{Aochi}.
               } 
The field equations 
derived from (\ref{L:vectorFroissart}) are
\begin{align}  
   & \partial^\mu (\partial_\mu Y_\nu - \partial_\nu Y_\mu )
                 - \varepsilon Y_{\ast \nu}=0, 
                           \label{Eq:Ym}
   \\
   & \partial^\mu (\partial_\mu Y_{\ast \nu} - \partial_\nu Y_{\ast \mu} )
                 =0,  
                             \label{Eq:Y*m}
\end{align}
from which we also have 
\begin{align}  
 & \partial^\nu Y_{\ast \nu}=0, 
  \\
  &  \Box Y_{\ast \nu} =0.         \label{Eq:BoxY*m}
\end{align}
From  (\ref{Eq:Ym}) and (\ref{Eq:BoxY*m}) we obtain the desired 
equation for $Y_\mu$:
\begin{align}
    \Box \partial^\mu (\partial_\mu Y_\nu - \partial_\nu Y_\mu )=0. 
                \label{Eq:DipoleVectorAgain}
\end{align}

%
%
%
\subsection{gauge fixing}

To quantize the Lagrangian (\ref{L:vectorFroissart}) we need appropriate
gauge-fixing terms since the Lagrangian 
is invariant under the gauge transformation 
\begin{align}  
        Y_\mu \to Y_\mu + \partial _\mu \Lambda,
\end{align}
where $\Lambda$ is an arbitrary scalar function. Our gauge fixed 
Lagrangian is 
\begin{align}  
    \mathcal{L}_{\mathrm {vF0+GF}}
    &= \mathcal{L}_{\mathrm{ vF0}}
       + Y_{\ast}^{ \mu} \partial_\mu Y
       +\partial ^\mu Y_{\ast} Y_\mu 
       + \beta' Y_\ast Y, 
                     \label{L:vF+GF}
\end{align}
where $Y_\ast$ and $Y$ are scalar multiplier fields and $\beta'$ is 
a gauge-fixing parameter. 
The field equations derived from (\ref{L:vF+GF}) are 
\begin{align}  
 & \partial^\mu (\partial_\mu Y_\nu - \partial_\nu Y_\mu )
                - \varepsilon Y_{\ast \nu}  + \partial_\nu Y=0, 
                          \label{Eq:YmGF}
  \\
  & \partial^\mu (\partial_\mu Y_{\ast \nu} - \partial_\nu Y_{\ast \mu} )
                  + \partial_\nu Y_{\ast}=0, 
                            \label{Eq:Y*mGF}& 
  \\
  & \partial ^\mu Y_\mu = \beta' Y ,
                             \label{Eq:divY}
   \\
  & \partial ^\mu Y_{\ast \mu} = \beta' Y_\ast ,
                             \label{Eq:divY*}
\end{align}
which lead to higher derivative field equations for $Y_\mu$, 
\begin{align}
  \Box^2 Y_\nu + \left( \frac{1}{{\beta'}^2} -1 \right) 
                         \Box \, \partial_\nu \partial^\mu Y_\mu=0. 
\end{align}
The higher derivative of the field equations suggests 
higher pole propagators. In fact,  we have
\begin{align}  
  & \langle Y_\mu Y_\nu \rangle 
       \sim \frac{\varepsilon}{(p^2)^2}
              \left[ g_{\mu\nu} + ({\beta'}^2-1)\frac{p_\mu p_\mu}{p^2} \right].
              \label{Prop:YmYn}
\end{align}

\subsection{BRST symmetry}
Because of the higher derivative field equations, 
the Fock space of the quantum theory 
derived from (\ref{L:vF+GF}) is not positive definite. 
We must remove these unphysical modes from the theory. This was done 
for the scalar Froissart model by introducing BRST symmetry \cite{Kashiwa, Sakoda}.
We imitate here again  the Froissart model with BRST symmetry. 

We introduce vector FP ghosts $K_\mu$ and $K_{\ast \mu}$, together with 
scalar FP ghosts $K$ and $K_\ast$, and define our Lagrangian by
\begin{align}  
 \mathcal{L}_\mathrm {vF}
      = &   -\frac{1}{2} (\partial^\mu Y_\ast^\nu - \partial^\nu Y_\ast^\mu )
                    (\partial_\mu Y_\nu - \partial_\nu Y_\mu )
        - \frac{\varepsilon}{2} Y_\ast^\mu Y_{\ast\mu} 
       %
         +\partial ^\mu Y_{\ast} Y_\mu 
         + Y_{\ast}^{ \mu} \partial_\mu Y
          + \beta' Y_\ast Y
       \notag \\
       &
       - \frac{i}{2} (\partial^\mu K_\ast^\nu - \partial^\nu K_\ast^\mu )
                    (\partial_\mu K_\nu - \partial_\nu K_\mu )
       +i \partial ^\mu K_{\ast} K_\mu 
         +i K_{\ast}^{ \mu} \partial_\mu K
          + i\beta' K_\ast K .
          \label{L:BRSTvF}
\end{align}
Note that the first term on the second line is invariant under the 
``gauge transformations" 
$K_\mu \to K_\mu + \partial _\mu \theta$ and 
$K_{\ast \mu} \to K_{\ast \mu} + \partial _\mu \theta_\ast$ 
where $\theta$ and  $\theta_\ast$ are arbitrary Grassmann odd functions. 
The remaining terms on the second line 
are the gauge-fixing terms for the gauge freedom; 
$K_\ast $ and $K$ play the role of the multiplier fields. 
These gauge-fixing terms are necessary for the vector FP ghosts to have  
propagators. 

 %
The Lagrangian (\ref{L:BRSTvF}) is invariant under the BRST transformations,  
\begin{align}  
    &
    \brst Y_\mu = K_\mu, \qquad \brst K_\mu =0, 
         \qquad \brst K_{\ast \mu} = iY_{\ast \mu}, 
                   \qquad \brst Y_{\ast \mu}=0,
    \notag \\
    &
    \brst Y= K, \qquad \brst K=0, \qquad \brst K_\ast = iY_\ast, 
                   \qquad \brst Y_\ast =0,    
                \label{BRST:vF}                   
\end{align}
which clearly satisfy the off-shell nilpotency $\brst^2=0$. 
The BRST invariance of (\ref{L:BRSTvF}) is  easlily confirmed 
when we rewrite (\ref{L:BRSTvF}) as
\begin{align}  
   \mathcal{L}_\mathrm {vF}
   = i\brst \Bigl[
                   \partial^\mu K_\ast^\nu 
                           (\partial_\mu Y_\nu - \partial_\nu Y_\mu )
                + \frac{\varepsilon}{2} K_\ast^\mu Y_{\ast \mu}
                - \partial^\mu K_\ast Y_\mu 
                - K_{\ast}^\mu \partial_\mu Y
                - \beta' K_\ast Y 
            \Bigr]
\end{align}
The corresponding 
BRST charge can be expressed by 
\begin{align}
   Q_\mathrm{ B(vF)}
      =\int \left[ 
                 Y_{\ast \mu} \overleftrightarrow{\partial _0} K^\mu
                 +(1-\beta')(  Y_{\ast 0} K - Y_\ast K_0)
            \right]  d^{D-1} x.
                \label{QB:vF}
\end{align}
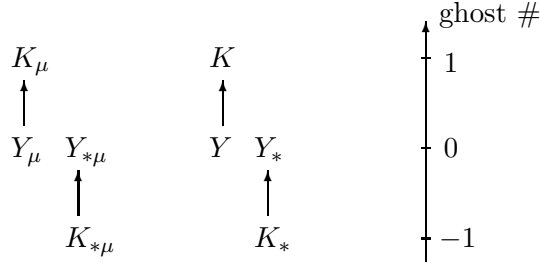
\begin{figure}
  \centering
  \setlength{\unitlength}{0.6mm}
  \begin{picture}(110,57)(0,-1)
    {\thinlines 
      \put(100,-1){\vector(0,1){53}}
         \put(103,52){ghost \#} 
          \put(104,41){$1$}    \put(99,44){\line(1,0){2}}
          \put(104,22){$0$}    \put(99,24){\line(1,0){2}}
          \put(103,2){$-1$}    \put(99, 4){\line(1,0){2}}
    }
     \put(8,22){$Y_{\mu}$} 
       \put(20,22){$Y_{\ast \mu}$} 
                      \put(52,22){$Y$}
                         \put(62,22){$Y_\ast$}
     \put(8,42){$K_\mu$}
       \put(20, 2){$K_{\ast \mu}$}
                      \put(52,42){$K$} 
                         \put(62, 2){$K_{\ast}$} 
     \put(11,29){\vector(0,1){10}}
       \put(23, 9){\vector(0,1){10}}
                      \put(55,29){\vector(0,1){10}}
                         \put(65,9){\vector(0,1){10}}
  \end{picture}
  \caption{
         Field contents and  BRST transformations of the BRST symmetric 
         vector-Froissart model.
         The arrows represent the directions of the BRST transformations. 
         }\label{fig:2}
\end{figure}
Figure \ref{fig:2} shows the field contents  and their BRST transformations of 
the quantized 
vector-Froissart model. All of the unphysical modes are removed 
by Kugo--Ojima's quartet mechanism;  any physical states satisfying 
$Q_\mathrm {B(vF)} |\mathrm{phys}\rangle=0$ are zero-normed states.

%
\section{gaugeon formalism}
\subsection{Lagrangian and field equations}
Combining the Lagrangians of Kimura's theory (\ref{L:Kimura}) and the  
massless vector-Froissart model (\ref{L:BRSTvF}), we present the 
Lagrangian of the gaugeon formalism for the two-form gauge field $B_{\mu\nu}$: 
\begin{align}  
  \mathcal{L} =& \mathcal{L}_\mathrm{K}(\alpha=0, \beta)
                + \mathcal{L}_\mathrm{vF}(\beta'=\beta)
                +\frac{\varepsilon}{2}Y_\ast^\mu Y_{\ast\mu}
                -\frac{\varepsilon}{2}
                        (Y_{\ast}^\mu + a B^\mu )(Y_{\ast\mu} + aB_\mu)
     \label{L:K+vF}
     \\
     =&          
     \frac{1}{12} F^{\lambda\mu\nu}F_{\lambda\mu\nu}
            -\partial ^\mu B^\nu B_{\mu\nu}
            + B^\mu \partial_\mu \eta 
            + \partial^\mu \phi_\ast \partial_\mu \phi 
      \notag \\
          & - \frac{\varepsilon}{2} 
                    (Y_{\ast}^\mu + a B^\mu )(Y_{\ast\mu} + aB_\mu)
      \notag \\ 
      &
          -\partial^\mu Y_\ast^\nu
                       (\partial_\mu Y_\nu - \partial_\nu Y_\mu )
           + Y_{\ast}^{ \mu} \partial_\mu Y
           +\partial ^\mu Y_{\ast} Y_\mu 
           + \beta Y_\ast Y
       \notag \\ &
        - i\partial^\mu c_{\ast}^{\nu}
                           (\partial_\mu c_\nu -\partial_\nu c_\mu)
            +i c_{\ast}^{\, \mu} \partial _\mu d 
            +i\partial^\mu d_\ast c_\mu 
            + i\beta d_\ast d 
       \notag \\
       &
       - i\partial^\mu K_\ast^\nu 
                    (\partial_\mu K_\nu - \partial_\nu K_\mu )
         +i K_{\ast}^{ \mu} \partial_\mu K
         +i \partial ^\mu K_{\ast} K_\mu 
          + i\beta K_\ast K ,
     \label{L:GaugeonFormalism}
\end{align}
where $a$ is a real parameter. 
The third and fourth terms of the right-hand-side of (\ref{L:K+vF}) 
show that the term 
$-(\varepsilon/2)Y_\ast^\mu Y_{\ast \mu}$ 
in $\mathcal{L}_\mathrm{vF}$ 
has been  replaced by 
$-(\varepsilon/2)(Y_\ast^\mu +aB^\mu)(Y_{\ast \mu}+aB_\mu)$. 
As seen later, 
the gauge-fixing parameter $\alpha$ of Kimura's theory (\ref{L:Kimura}) 
can be identified 
through the parameter $a$ as
\begin{align}  
         \alpha = \varepsilon a^2.
              \label{a:alpha=ea^2}
\end{align}
The field equations derived from (\ref{L:GaugeonFormalism}) are 
\begin{align}  
  & 
  \partial ^\lambda F_{\lambda \mu \nu} 
               + \partial_\mu B_\nu - \partial_\nu B_\mu =0, 
  \notag \\ 
  &
  \partial^\mu B_{\mu\nu} + \partial_\nu \eta 
           -\varepsilon a (Y_{\ast \nu} + a B_\nu) =0,
  \notag \\
  &
   \partial^\mu B_\mu =   \Box \phi = \Box \phi_\ast =0, 
  \notag \\
  &
  \partial^\mu (\partial _\mu Y_\nu - \partial_\nu Y_\mu ) 
          - \varepsilon (Y_{\ast \nu} + a B_\nu) + \partial _\nu Y =0,
              \qquad \partial^\mu Y_\mu = \beta Y
  \notag \\ 
  &
  \partial^\mu (\partial _{\mu} Y_{\ast \nu} - \partial_\nu Y_{\ast \mu} ) 
          + \partial _\nu Y_\ast =0,
            \qquad \partial^\mu Y_{\ast \mu}=\beta Y_\ast,
\intertext{for bosonic 
fields, and}  
  &
  \partial^\mu (\partial _{\mu} c_\nu - \partial_\nu c_{ \mu} ) 
          + \partial_\nu d =0, \qquad \partial^\mu c_\mu = \beta d, 
  \notag \\
  &
  \partial^\mu (\partial _{\mu} c_{\ast \nu}  - \partial_\nu c_{\ast \mu} ) 
          + \partial_\nu d_\ast  =0, 
                                \qquad \partial^\mu c_{\ast \mu}= \beta d_\ast,
  \notag \\
  &
  \partial^\mu (\partial _{\mu} K_\nu - \partial_\nu K_{ \mu} ) 
          + \partial_\nu K =0, \qquad \partial^\mu K_\mu = \beta K, 
  \notag \\
  &
  \partial^\mu (\partial _{\mu} K_{\ast \nu}  - \partial_\nu K_{\ast \mu} ) 
          + \partial_\nu K_\ast  =0, 
                                \qquad \partial^\mu K_{\ast \mu}= \beta K_\ast,
\end{align}
for fermionic 
fields. 
We emphasize here that 
we have chosen the gauge-fixing parameter 
$\beta'$  of  the 
vector-Froissart fields $Y_\mu$ and $Y_{\ast \mu}$ as 
$\beta'=\beta$. 
As a result, 
four  pairs of FP ghosts 
($c_\mu$, $d$), ($c_{\ast \mu}$, $d_\ast$), 
($K_\mu$, $K$) and ($K_{\ast \mu}$, $K_\ast$)
, as well as ($Y_{\ast \mu}$, $Y_\ast$), 
satisfy the same field equations. 
%
%
%
\subsection{BRST symmetry}
\begin{figure}
  \centering
  \setlength{\unitlength}{0.6mm}
  \begin{picture}(220,110)(10,-12)
    {\thinlines 
      \put(222,-3){\vector(0,1){93}}
         \put(225,90){ghost \#} 
          \put(226,79){$2$}    \put(221,82){\line(1,0){2}}
          \put(226,60){$1$}    \put(221,62){\line(1,0){2}}
          \put(226,40){$0$}    \put(221,42){\line(1,0){2}}
          \put(225,20){$-1$}   \put(221,22){\line(1,0){2}}
          \put(225,0){$-2$}    \put(221, 2){\line(1,0){2}}
      }
   %
   %
     \put(0,40){$B_{\mu\nu}$ ($B_{\mu\nu}$)}
          \put(37,40){$B_\mu$ ($\beta_{\mu}$)}
                 \put(71,40){$\eta$ ($\varphi$)}
          \put(37,60){$c_\mu$ ($\rho_\mu$)}
          \put(37,20){$c_{\ast \mu}$ ($\bar \rho_\mu$)}
                 \put(71,60){$d$ ($\chi$)}
                 \put(71,20){$d_{\ast}$ ($\bar \chi$)}
                 \put(71,80){$\phi$ ($\sigma$)}
                 \put(71,0){$\phi_{\ast}$ ($\bar \sigma$)}
     \put(07,46){\vector(2,1){22}}
     \put(46,66){\vector(2,1){22}}
          \put(42,26){\vector(0,1){10}}
                 \put(75,46){\vector(0,1){10}}
                 \put(75, 6){\vector(0,1){10}}
  %
  %
       \put(108,40){$Y_{\mu}$ ($Y_\mu$)} 
       \put(138,40){$Y_{\ast \mu}$ ($Y^\star_\mu$)} 
                      \put(177,40){$Y$}
                         \put(192,40){$Y_\ast$}
                       \put(182,80){($Z$)}
     \put(108,60){$K_\mu$($K_\mu$)}
       \put(138,20){$K_{\ast \mu}$($K^\star _\mu$)}
                      \put(177,60){$K$} 
                         \put(192,20){$K_{\ast}$} 
                       \put(182,0){($Z^\star$)}
     \put(116,46){\vector(0,1){10}}
       \put(146,26){\vector(0,1){10}}
                      \put(179,46){\vector(0,1){10}}
                         \put(194,26){\vector(0,1){10}}
  %
    \put(8,-15){standard formalism sector}
    \put(137,-15){gaugeon sector}
  %
 \end{picture}
 \caption{
         Field contents and  BRST transformations of the gaugeon formalism. 
         The arrows show the BRST transformations. 
         The fields in the parentheses represent 
         corresponding fields of Refs.\cite{UpadhyayPanigrahi, Dwivedi}; 
         there are no counterparts of our fields $Y$, $Y_\ast$, $K$, and $K_\ast$. 
         Instead, two BRST-singlet fields $Z$ and $Z^\star$ were introduced  
         as ghosts for ghosts
         in \cite{UpadhyayPanigrahi, Dwivedi}.  
         }\label{fig:3}
\end{figure}
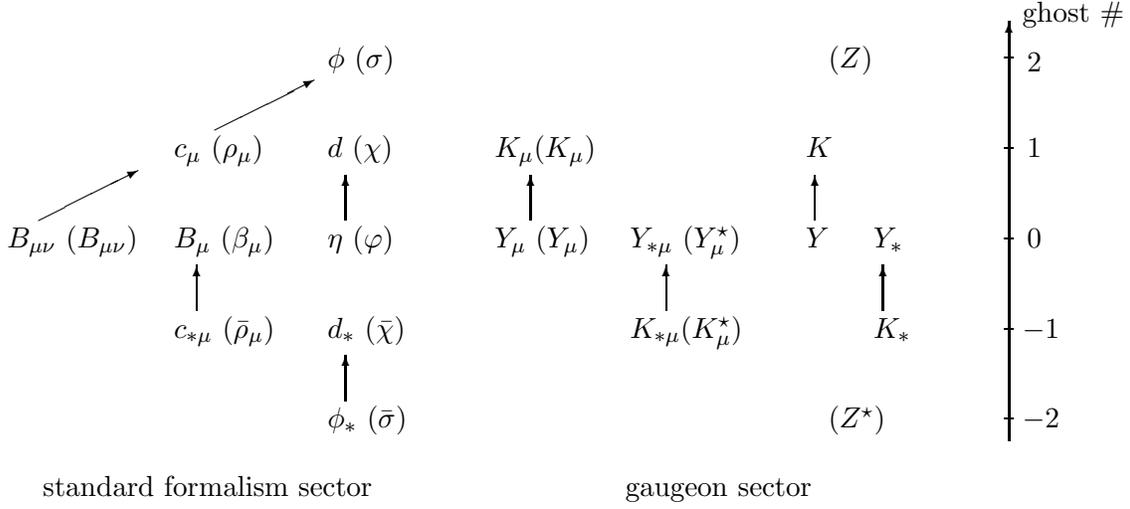

The Lagrangian (\ref{L:GaugeonFormalism}) is invariant under the 
BRST transformations which are defined by 
\begin{align}
  &\brst B_{\mu\nu}= \partial_\mu c_\nu -  \partial_\nu c_\mu,
    \qquad 
    \brst c_\mu= -i \partial_\mu  \phi, 
    \notag \\
  & \brst c_{\ast \mu}= iB_\mu, 
      \qquad 
        \brst \phi_\ast = d_\ast, 
          \qquad
             \brst \eta = d, 
    \notag 
    \\%
  & \brst B_\mu= \brst d_\ast =\brst d = \brst \phi=0. 
    \label{BRST2:Kimura}
\intertext{for the fields of the standard formalism sector and}
    &
    \brst Y_\mu = K_\mu, \qquad \brst K_\mu =0, 
         \qquad \brst K_{\ast \mu} = iY_{\ast \mu}, 
                   \qquad \brst Y_{\ast \mu}=0,
    \notag \\
    &
    \brst Y= K, \qquad \brst K=0, \qquad \brst K_\ast = iY_\ast, 
                   \qquad \brst Y_\ast =0,    
                \label{BRST2:vF}                   
\end{align}
for the fields of gaugeon sector. 
Field contents and their BRST transformations are shown 
in Figure \ref{fig:3}. (Fields introduced in Refs.\cite{UpadhyayPanigrahi, Dwivedi} are 
also shown in the figure for comparison.) 
Because of the off-shell nilpotency $\brst^2=0$, 
the BRST invariance of the Lagrangian is easily understood when 
we rewrite  (\ref{L:GaugeonFormalism}) as 
\begin{align}  
  \mathcal L =  & \frac{1}{12}  F^{\lambda\mu\nu}F_{\lambda\mu\nu}
      + i \brst \Bigl [\,  
             \partial^\mu c_\ast^\nu
                            (\partial_\mu B_\nu - \partial_\nu B_\mu)
               \notag \\
               &- c_\ast^\mu \partial _\mu \eta + \partial^\mu \phi_\ast c _\mu  
                                                      -\beta d_\ast \eta 
                + \frac{\varepsilon}{2} (K_\ast^\mu + a c_\ast^\mu) 
                                           ( Y_{\ast \mu } + a B_\mu) 
               \notag \\
               &
               + \partial^\mu K_\ast^\nu
                            (\partial_\mu Y_\nu - \partial_\nu Y_\mu)
               - K_\ast^\mu \partial _\mu Y + \partial^\mu K_\ast Y _\mu  
                                                     -\beta K_\ast Y
               \, \Bigr]. 
\end{align}
The corresponding  BRST charge $Q_{\rm B}$ 
can be written as 
\begin{align}
  Q_\mathrm {B} = \int \Big[\,
                    & B^\lambda \overleftrightarrow{\partial_0} c_\lambda
                   + d_\ast \overleftrightarrow{\partial_0} \phi
                   + (1-\beta) B_0 d
                   \notag \\
                   + & Y_{\ast}^\lambda \overleftrightarrow{\partial _0} K_\lambda
                   +(1-\beta)(  Y_{\ast 0} K - Y_\ast K_0)
            \, \Big]  d^{D-1} x.
                \label{QB:GaugeonFormalism}
\end{align}
With the help of the charge, we can define the physical subspace of the 
Fock space by 
\begin{align}  
   \mathcal V _{\rm phys} = \ker Q_{\rm B}
               = \{ |\Phi \rangle; \, Q_{\rm B} |\Phi \rangle =0  \}. 
   \label{V:phys}
\end{align}
%
%
%
\subsection{$q$-number gauge transformations}
The Lagrangian (\ref{L:GaugeonFormalism}) permits the  
$q$-number gauge transformation where we vary the gauge-fixing parameter $a$. 
Under the field redefinitions 
\begin{align}  
   & 
   \hat B_{\mu\nu}= B_{\mu\nu} 
                    + \tau (\partial_\mu Y_\nu - \partial_\nu Y_\mu), 
     \quad 
     \hat c_\mu = \hat c_\mu + \tau K_\mu, 
  \notag \\ &
     \hat Y_{\ast \mu} = Y_{\ast \mu} - \tau B_\mu, 
     \quad
     \hat K_{\ast \mu} = K_{\ast \mu} - \tau c_{\ast \mu},
  \notag \\ &
     \hat B_\mu = B_\mu, \quad \hat Y_\mu = Y_\mu, 
     \quad 
     \hat c_{\ast \mu}= c_{\ast \mu}, \quad \hat K_\mu = K_\mu, 
  \notag \\ &
     \hat \eta = \eta + \tau Y , \quad \hat d= d + \tau K, 
  \notag \\ &
     \hat Y_\ast = Y_\ast,
     \quad 
     \hat K_{\ast }= K_\ast - \tau d_\ast, 
  \notag \\ &
     \hat Y= Y, \quad \hat d_\ast = d_\ast,  
     \quad \hat K=K,
     \quad \hat \phi = \phi, 
     \quad \hat \phi_\ast = \phi_\ast, 
     \label{T:q-number}
\end{align}
with  $\tau$ being  a real parameter, 
the Lagrangian  (\ref{L:GaugeonFormalism}) 
becomes 
\begin{align}  
  \mathcal L (\varPhi_A; a, \beta ) = \mathcal L (\hat \varPhi_A; \hat a, \beta),
  \label{FormInvariance:L}
\end{align}
where $\varPhi_A$ collectively represents all fields and 
$\hat a$ is defined by 
\begin{align}  
       \hat a = a + \tau.
\end{align}
The {\it form invariance} (\ref{FormInvariance:L}) concludes 
that the field equations 
transform  gauge covariantly under the $q$-number gauge 
transformations (\ref{T:q-number}):   
$\hat \varPhi_A$ satisfies the same field equation as $\varPhi_A$ if the parameter 
$a$ replaced by $\hat a$. 

It should be noted that the $q$-number gauge transformations (\ref{T:q-number}) 
commute with 
the BRST transformations 
(\ref{BRST2:Kimura}) and (\ref{BRST2:vF}). 
As a result, the BRST charge is invariant under 
the $q$-number gauge transformations: 
\begin{align}  
    \hat Q_\mathrm{B} =  Q_\mathrm{ B}. 
\end{align}
The physical subspace $\mathcal V_\mathrm{phys}$ is, therefore, also 
invariant under the $q$-number gauge transformation: 
\begin{align}  
    \hat{\mathcal{V}}_\mathrm{phys}= \mathcal{V}_\mathrm{phys}. 
\end{align}

\subsection{gauge structure of the Fock space}
In addition to the BRST 
 charge (\ref{QB:GaugeonFormalism}), 
the Lagrangian (\ref{L:GaugeonFormalism}) has several 
conserved BRST-like charges. 
We focus on here 
the following three charges:
\begin{align}  
  Q_\mathrm{B(K)} &= \int 
                  \left[
                        B^\mu \overleftrightarrow{\partial_0} c_\mu
                        + d_\ast \overleftrightarrow{\partial_0} \phi
                        + (1-\beta) B_0 d
                   \right] d^{D-1} x ,
                  \label{QB2:Kimura}
  \\
  Q_\mathrm{B(vF)} 
      & =\int \left[ 
                 Y_{\ast}^{ \mu} \overleftrightarrow{\partial _0} K_\mu
                 +(1-\beta')( Y_{\ast 0} K - Y_\ast K_0)
            \right]  d^{D-1} x,
            \label{QB2:vF}
  \\
  \tilde Q_\mathrm{B}
       & =\int \left[ 
                 B^\mu \overleftrightarrow{\partial _0} K_\mu
                 +(1-\beta') B_0 K
            \right]  d^{D-1} x, 
      \label{QB2:tilde}
\end{align}
which are  nilpotent and anticommuting with each other:
\begin{align} 
   & Q_\mathrm{B(K)}^{\,2}= Q_\mathrm{B(vF)}^{\,2} = Q_\mathrm{B}^{\,2} =0, 
   \notag \\
   & \{ Q_\mathrm{B(K)}, Q_\mathrm{B(vF)} \}
   = \{ Q_\mathrm{B(vF)}, \tilde Q_\mathrm{(B)} \}
   = \{ \tilde Q_\mathrm{B}, Q_\mathrm{B(K)} \}
     =0.
\end{align}
The 
$Q_\mathrm{B(K)}$ 
generates the BRST transformation 
(\ref{BRST2:Kimura}) acting only on the fields of the standard formalism sector, 
while 
$Q_\mathrm{B(vF)}$ 
generates the transformation (\ref{BRST2:vF}) 
acting only on the fields of the gaugeon sector. The total BRST charge 
(\ref{QB:GaugeonFormalism}) can be expressed as 
\begin{align}
         Q_\mathrm{B}=Q_\mathrm{B(K)}+Q_\mathrm{B(vF)}.
\end{align} 
The charge $\tilde Q_\mathrm{B}$ (\ref{QB2:tilde}) generates the  transformation 
$\tbrst$: 
\begin{align}  
  & \tbrst B_{\mu\nu} = \partial_\mu K_\nu - \partial_\nu K_\nu, 
  \qquad 
   \tbrst K_{\ast \mu} = i B_\mu, 
  \qquad 
  \tbrst \eta = K , 
  \notag \\
  & \tbrst (\text{other fields}) =0.
\end{align}

In (\ref{V:phys}) we have defined the physical subspace $\mathcal V_\mathrm{phys}$
using the charge 
$Q_\mathrm{B}$. 
Instead, we may consider another subspace by 
\begin{align}  
   \mathcal V ^{(a)}_\mathrm{phys}=\ker Q_\mathrm{B(K)} \cap \ker Q_\mathrm{B(vF)}
                  = \{ |\Phi \rangle ; \, Q_\mathrm{B(K)}|\Phi \rangle
                                = Q_\mathrm{B(vF)} |\Phi \rangle =0 \}.
\end{align}
The condition $Q_\mathrm{B(K)} |\Phi \rangle =0$ removes the unphysical 
modes included in  the standard formalism sector, while the 
condition $Q_\mathrm{B(vF)} |\Phi \rangle =0$ removes the modes of the 
gaugeon sector. As is easily seen, the space $\mathcal V^{(a)}_\mathrm{phys}$ 
is a subspace of $\mathcal V _\mathrm{phys}$: 
\begin{align}  
    \mathcal V^{(a)}_\mathrm{phys} \subset \mathcal V _\mathrm{phys}.
\end{align}
We have attached the index $(a)$ to  
$\mathcal V^{(a)}_\mathrm{phys}$ 
to emphasize that its definition depends on the gauge-fixing parameter $a$. 
In fact, under the $q$-number gauge transformation (\ref{T:q-number}), 
the BRST charges 
$Q_\mathrm{B(K)}$ 
and 
$Q_\mathrm{B(vF)}$ 
transform as 
\begin{align}  
   &\hat Q_\mathrm{B(K)} = Q_\mathrm{B(K)} + \tau \tilde Q_\mathrm{B}, 
   \notag \\
   &\hat Q_\mathrm{B(vF)} = Q_\mathrm{B(vF)} - \tau \tilde Q_\mathrm{B}, 
   \label{QB:hatted}
\end{align}
while their sum $Q_\mathrm{B}$ remains invariant. 

Let us define a subspace  $\mathcal V^{(a)}$ of the total Fock space by
\begin{align}  
           \mathcal V^{(a)} = \ker Q_\mathrm{B(vF)}.  
\end{align}
This space can be identified with the total Fock space of the standard 
formalism in the $\alpha=\varepsilon a^2$ gauge. 
We can understand this by rewrite the Lagrangian (\ref{L:GaugeonFormalism}) 
as, 
\begin{align}  
  \mathcal L = \mathcal L_\mathrm{K}(\alpha=\varepsilon a^2)
            +i \{ Q_\mathrm{B}, \varTheta \} ,
     \label{Manifestation:a-gauge}
\end{align}
where 
$\varTheta$ being  given by
\begin{align}
  \varTheta = 
                \frac{\varepsilon}{2} K_\ast^\mu  
                                           ( Y_{\ast \mu } + 2a B_\mu) 
                +\partial^\mu K_\ast^\nu 
                            (\partial_\mu Y_\nu - \partial_\nu Y_\mu)
               - K_\ast^\mu \partial _\mu Y + \partial^\mu K_\ast Y _\mu  
                                                     -\beta K_\ast Y.  
\end{align}
The first term of (\ref{Manifestation:a-gauge}) corresponds to the 
Lagrangian of the standard formalism (\ref{L:Kimura}). The second term 
becomes null-operator in the subspace $\mathcal V^{(a)}$. Namely, we can 
ignore the second term of  (\ref{Manifestation:a-gauge}) in  $\mathcal V^{(a)}$. 

We emphasize that the same  arguments hold  if we start from the $q$-number 
transformed charges (\ref{QB:hatted}) 
rather than $Q_\mathrm{B(K)}$ and $Q_\mathrm{B(vF)}$. 
We define the subspaces 
$\mathcal V^{(a+\tau)}$ and  
$\mathcal V^{(a+\tau)}_\mathrm{phys}$ by 
\begin{align}  
  &\mathcal V^{(a+\tau)} = \ker \hat Q_\mathrm{B(vF)}, 
  \notag \\
  &\mathcal V^{(a+\tau)}_\mathrm{phys} 
          = \ker \hat Q_\mathrm{B(K)} \cap 
           \ker \hat Q_\mathrm{B(vF)}. 
\end{align}
The space $\mathcal V^{(a+\tau)}$ can be identified with the Fock space of the 
standard formalism in the $\alpha =\varepsilon (a+\tau)^2$ gauge, and 
$\mathcal V^{(a+\tau)}_\mathrm{phys}$ 
corresponds to its physical subspace. 
Thus  various Fock spaces of the standard formalism in different gauges 
are embedded in the single Fock space of the present theory.


\section{Summary and comments}
We have presented the BRST symmetric gaugeon formalism for the 
two-form gauge theory. For this purpose, we have covariantly quantized 
the massless vector-Froissart model (a dipole vector gauge theory), 
as  vector gaugeon fields of our theory. 
Since this model has gauge invariance at the classical level, 
we have first considered  gauge-fixing for the model; the necessity 
of the gauge-fixing for the gaugeon fields  was overlooked in the previous 
literature \cite{UpadhyayPanigrahi,Dwivedi}. 
Using three kinds of BRST charges as well as the total BRST charge,  
we have shown that our total Fock space 
contains the subspaces which are identified with 
the Fock spaces of the standard formalism in various gauges. 

In the following, we add some comments. 

\subsection{Type II theory}
In the  theory presented in the last section, 
we can change the value of the gauge-fixing parameter $a$ by the 
$q$-number gauge transformation (\ref{T:q-number}). 
The gauge-fixing parameter $\alpha$ of the standard 
formalism (\ref{L:Kimura}) is identified with 
$\alpha= \varepsilon a^2$ ($\varepsilon=\pm 1$). 
This means that 
we cannot change the sign of the parameter of the 
standard parameter $\alpha$ by the $q$-number transformation. 
The situation is analogous to  Type I gaugeon theory for QED \cite{Yokoyama74b}. 
There are two types of gaugeon theories, Type I and Type II. 
The gauge-fixing parameter $a$ 
can be shifted as $\hat a= a+ \tau$ by the $q$-number gauge transformation 
in  both theories. 
The standard gauge-fixing parameter $\alpha$ is expressed as 
$\alpha=\varepsilon a^2$ in Type I theory, and  $\alpha=a$ in Type II theory; 
the sign of $\alpha$ can be changed in Type II theory. 
We comment here that Type II theory can also be formulated for the 
two-form gauge fields. 

We consider the Lagrangian, rather than (\ref{L:K+vF}),
\begin{align}  
  \mathcal{L}_\mathrm{II} 
      & = \mathcal L_\mathrm{K}(\alpha=a;\beta)
          + \mathcal{L}_\mathrm{vF}(\beta'=\beta) 
          +\frac{\varepsilon}{2}Y_\ast^\mu Y_{\ast \mu} 
          -\frac{1}{2} Y_\ast^\mu B_\mu . 
     \label{L:typeII}
\end{align}
Under the $q$-number transformation (\ref{T:q-number}), this Lagrangian 
is also form invariant:
\begin{align}
   \mathcal L_\mathrm{II} (\varPhi_A; a, \beta) = 
   \mathcal L_\mathrm{II} (\hat \varPhi_A; \hat a, \beta) 
\end{align}
with $\hat a=a+\tau$. The standard gauge-fixing parameter $\alpha$ 
can be identified with 
\begin{align}
    \alpha = a
\end{align}
in the present case, thus we can change the parameter $\alpha$ quite 
freely 
without any  limitation for the sign of $\alpha$. 

The Lagrangian (\ref{L:typeII}) is also invariant under all of the 
transformations corresponding to the BRST charges 
(\ref{QB:GaugeonFormalism}), 
(\ref{QB2:Kimura}), 
(\ref{QB2:vF}), and
(\ref{QB2:tilde}). 
Thus, the similar arguments to those in the last section 
on the gauge structure of the Fock space are also available. 
For example, 
$\mathcal V^{(a)}=\ker Q_\mathrm{B(vF)}$  
[$\mathcal V^{(a+\tau)}=\ker \hat Q_\mathrm{B(vF)}$ ] 
is identified with 
the Fock space of the standard formalism in $\alpha=a$ [$\alpha=a+\tau$] gauge.

\subsection{gaugeons for gaugeons}
The standard formalism (\ref{L:Kimura}) has two gauge-fixing parameters 
$\alpha$ and $\beta$. As seen in the last section (and in the last subsection), 
the value of the parameter $\alpha$ can be changed by the 
$q$-number gauge transformation (\ref{T:q-number}), while the value of 
$\beta$ cannot. 
One might attempt to find  a $q$-number transformation which 
can change the value of the parameter $\beta$, 
the  gauge-fixing parameter for the  FP ghosts 
$c_\mu$ and $c_{\ast \mu}$. 
Let us consider this possibility here. 

To introduce the $q$-number gauge transformation for the vector FP ghosts, 
we  would need ghost-number $\pm 1$ gaugeon fields (gaugeons for ghosts) 
and their FP ghosts (ghosts for gaugeons for ghosts); 
the FP ghosts have $\pm 2$ ghost numbers and thus  might be identified with the 
fields $Z$ and $Z^\star$ introduced in Refs.\cite{UpadhyayPanigrahi, Dwivedi}. 
Furthermore, remembering that $\beta\,(=\beta')$ is  a gauge-fixing parameter also 
for the gaugeon fields $Y_\mu$ and $Y_{\ast \mu}$, we would need zero-ghost-number 
gaugeons (gaugeons for gaugeons) too, 
and their FP ghosts (ghosts for gaugeons for gaugeons). 
An  early attempt of this program is seen in Ref. \cite{AochiMaster}.

%
%
\begin{thebibliography}{99}
\bibitem{Nakanishi}N.~Nakanishi, Prog. Theor. Phys. Suppl.%
       {{\bf 51}, 1 (1972).} 
\bibitem{Kugo78a}T.~Kugo and I.~Ojima, Phys. Lett. B %
        {{\bf 73}, 459 (1978).}
\bibitem{Kugo78b}T.~Kugo and I.~Ojima, Prog. Theor. Phys. %
        {{\bf 60}, 1869 (1978).} 
\bibitem{Kugo79}T.~Kugo and I.~Ojima, Prog. Theor. Phys. Suppl. %
        {{\bf 66}, 1 (1979).} 
\bibitem{Kugo89}T.~Kugo, {\it Quantum Theory of Gauge Field I, II} (Baifukan, Tokyo, 1989), [in Japanese].
%
\bibitem{Yokoyama74a}K.~Yokoyama, Prog. Theor. Phys. %
   {{\bf 51}, 1956 (1974).} 
\bibitem{Yokoyama74b}K.~Yokoyama and R.~Kubo, Prog. Theor. Phys. %
   {{\bf 52}, 290 (1974).}  
\bibitem{Izawa}K.~Izawa, Prog. Theor. Phys. %
   {{\bf 88}, 759 (1992).} 
\bibitem{Koseki}M.~Koseki, M.~Sato, and R.~Endo, Prog. Theor. Phys. %
    {\bf 90}, 1111 (1993). 
\bibitem{Endo}R.~Endo, Prog. Theor. Phys. %
   {{\bf 90}, 1121 (1993).}   
\bibitem{Saito}T. Saito, R. Endo, and H. Miura,  Prog. Theor. Exp. Phys. %
   {{\bf 2016}, 023B02 (2016).} 
%
\bibitem{Yokoyama78b}K.~Yokoyama, Prog. Theor. Phys. %
   {{\bf 59}, 1699 (1978).} 
\bibitem{Yokoyama78c}K.~Yokoyama, M.~Takeda, and M.~Monda, Prog. Theor. Phys. %
   {{\bf 60}, 927 (1978).}  
\bibitem{Yokoyama78d}K.~Yokoyama, Prog. Theor. Phys. %
   {{\bf 60}, 1167 (1978).} 
\bibitem{Yokoyama78e}K.~Yokoyama, Phys. Lett. B %
   {{\bf 79}, 79 (1978).} 
\bibitem{Yokoyama80}K.~Yokoyama, M.~Takeda, and M.~Monda, Prog. Theor. Phys. %
   {{\bf 64}, 1412 (1980).} 
\bibitem{Abe}M.~Abe, {\it The Symmetries of the Gauge-Covariant Canonical Formalism of Non-Abelian Gauge Theories}, Master Thesis, Kyoto University, 1985.
\bibitem{Koseki96}M.~Koseki, M.~Sato, and R.~Endo, Bull. of Yamagata Univ., Nat. Sci.%
       {{\bf 14}, 15 (1996)} (doi:10.15022/00003947)
    \lbrack 
    Errata: R. Endo, Bull. of Yamagata Univ., Nat. Sci. %
     {\bf 18}, 33 (2017) (doi:10.15022/00004109)
    \rbrack . 
\bibitem{Sakoda}S. Sakoda, Prog. Theor. Phys. %
    {\bf 117}, 745 (2007).
%
\bibitem{Yokoyama75a}K.~Yokoyama and R.~Kubo, Prog. Theor. Phys. %
   {{\bf 54}, 848 (1975).} 
\bibitem{Miura}H.~Miura and R.~Endo, Prog. Theor. Phys. %
     {{\bf 117}, 695 (2007).}
%
\bibitem{Yokoyama75b}K.~Yokoyama, S.~Yamagami, and R.~Kubo, Prog. Theor. Phys. %
    {{\bf 54}, 1532 (1975).} 
%
\bibitem{Nakawaki} Y.~Nakawaki, Prog. Theor. Phys. %
     {{\bf 98}, 1193 (1997).} 
%
\bibitem{Endo00}R.~Endo and M.~Koseki, Prog. Theor. Phys. %
     {{\bf 103}, 685 (2000).} 
%
\bibitem{Faizal12a}M.~Faizal, Commun. Theor. Phys. %
     {{\bf 57}, 637 (2012).} 
\bibitem{Faizal12b}M.~Faizal, Mod. Phys. Lett. %
     {{\bf A27}, 1250147 (2012).} 
%
\bibitem{Upadhyay14a}S.~Upadhyay, Ann. Phys. %
   {{\bf 344}, 290 (2014).} 
\bibitem{Upadhyay14b}S.~Upadhyay, Eur. Phys. J. C %
   {{\bf 74} 2737(2014).} 
%
\bibitem{UpadhyayPanigrahi}
   S. Upadhyay and P. K. Panigrahi, Nucl. Phys. B %





\end{document}